# The strain gradient elasticity via nonlocal considerations

T. Gortsas[1], D. G. Aggelis[2], D. Polyzos[1*]

[1]Department of Mechanical Engineering and Aeronautics, University of Patras, Patras, Greece

[2]Department of Mechanics of Materials and Constructions, Vrije Universiteit Brussel, Brussels, Belgium

[*]Corresponding Author, *E-mail address* polyzos@upatras.gr (Prof. Demosthenes Polyzos, University of Patras, Rio-Patras, GR265 04, Greece)

**Abstract**

Strain gradient elasticity and nonlocal elasticity are two enhanced elastic theories intensively used over the last fifty years to explain static and dynamic phenomena that classical elasticity fails to do. The nonlocal elastic theory has a clear differentiation from the classical case by considering stresses in a point of the continuum as an integral of all stresses defined in the treated elastic body. On the other hand, strain gradient elasticity is characterized as a non-classical theory because considers both potential and kinetic energy densities as not only functions of strains and velocities, respectively but also functions of their gradients. Although the two considerations seem to be completely different from each other, it is a common belief that strain gradient elasticity has a lot in common with nonlocal elasticity. The goal of the present work is to derive all the strain gradient elastic theories appearing so far in the literature via nonlocal definitions of the potential and kinetic energy densities. Such a derivation demonstrates the common roots of the two theories and explains the nature of the involved intrinsic parameters in strain gradient elastic theories. For the sake of simplicity and brevity, only one-dimensional wave propagation phenomena are considered.

## 1. Introduction

The classical theory of elasticity has been almost completely formed by Navier, Cauchy and Green during the first half of the nineteenth century, as it is recited by Love (1927) in the introduction of his book. However, before one tackle non-classical elastic theories, it would be useful to know the basic assumptions made by the classical elastic theory. These are (i) the kinematical assumption, (ii) the dynamical assumption and (iii) the constitutive assumption (Rogula, 1985).

The **kinematical assumption** idealizes the material body as a continuum consequence of structureless entities behaving like material points. The atomic structure of the matter is not taken into account and the continuum hypothesis saying that the matter of an elastic body is homogeneous and continuously distributed over its volume so that the smallest element cut from the body possesses the same specific physical properties as the body (Timoshenko and Goodier, 1970), is valid for any classical elastic medium. In other words, the material properties of the continuum, assigned at any point **x**, are taken as mean values over very small volume elements where the point **x** is centered. The size of those small volume elements is very large as it is compared to the atomic structure of the material and very small compared to any dimension of the considered deformable macrostructure. In a nutshell, the classical elastic material is a continuum with an infinite number of particles that move and interact only with their neighbors. The **dynamical assumption** imposes the following constraints on the aforementioned small volume elements corresponding to each material point **x** of the continuum at any time t. These are: (a) there are only contact interactions, which are represented via force vectors located at point **x**, (b) the force vectors are uniquely determined by the stress tensor and (c) the stress tensor is symmetric meaning that no net moment of forces appear from the interaction between particles. Finally, the **constitutive assumption** postulates that the symmetric stress tensor defined at any material point **x** can be entirely determined from the deformation tensor at the same point. All these assumptions render the classical theory of elasticity strictly local and unable to distinguish the difference between small and large or as it is pointed out by Exadaktylos and Vardoulakis (2001) *"...this in turn leads to the undesirable result that a 10 cm slab behaves the same as a 10 μm film and there is no difference between a microcrack and a geological fault."*

An elastic theory is characterized as enhanced or nonclassical elastic theory if it makes more realistic assumptions than those of classical theory and introduces new degrees of freedom and/or new parameters that capture static or dynamic phenomena that classical elasticity is not able to do. Depending on the type of enhancement of the classical theory of elasticity, many such theories have been proposed during mainly the second half of the 20[th] century with the most known being those of Cosserat brothers (Cosserat and Cosserat, 1909), couple stresses (Truesdell and Toupin,1960; Mindlin and Tiersten, 1962; Koiter, 1964), multipolar elasticity (Green and Rivlin, 1964), strain gradient elasticity (Toupin, 1962, 1964; Mindlin, 1964, 1965; Germain, 1973), strain gradient elasticity with surface energy (Casal 1961; Vardulakis and Sulem, 1995), micromorphic, microstretch, micropolar elasticity (Eringen and Suhubi, 1964; Eringen, 1966; Eringen, 1990; Eringen, 1999), nonlocal elasticity (Kroner, 1967; Eringen and Edelen, 1972; Eringen, 1972), stress gradient elasticity (Eringen 1983, Forest and Sab, 2012), elasticity with surface stresses (Gurtin and Murdoch, 1975), peridynamics (Silling, 2000) and nonlocal strain gradient elasticity (Lim *et al*., 2015). Comprehensive reviews of those theories as they compared to each other can be found, among others, in the works of Rogula (1982), Jirásek (2004), Tekoglu and Onck (2008), Forest (2009), Borino and Polizzotto (2014), Polizzotto (2014, 2017), Broese *et al*., 2016, Maugin (2017) and Shaat *et al*. (2020).

Undoubtedly, the three nonclassical theories that have attracted the most attention among the researchers are those of micropolar, strain gradient, and nonlocal elasticity with the last two being the most discussed and compared to each other since their dynamic behavior is described by opposite types of equations. The equation of motion in strain gradient elasticity (SGE) is a set of partial differential equations while in nonlocal elasticity (NLE) a set of integro-differential ones. This is because, at any point of the domain of interest, the total stresses in SGE are determined from the strain tensor and its higher-order gradients defined at $x$, while in NLE the stress tensor is recovered by integrating the classical Cauchy tensor in the entire domain of interest. For this reason, the SGE belongs to the so-called weakly nonlocal theories while the NLE is known as strongly nonlocal theory. Fruitful discussions on that can be found in Rogula (1985), Jirásek (2004) and Bažant and Jirásek, M., (2002).

The departure point of the SGE theories goes back to the landmark papers of Toupin (1962) and Mindlin (1964). Mindlin motivated by his work in couple stresses theory, mentions in his work Mindlin (1963) that it would be interesting to explore one the consequences of considering the first and higher gradients of strain in the expressions of potential energy for solving linear elastic problems in materials with microstructure. One year later, with his landmark paper Mindlin (1964), he proposed a new linear elastodynamic theory by introducing a deformable elastic microstructure embedded in a Cauchy space and considering the potential energy density as a function of strains and first gradient of strains, while the kinetic energy density is represented via velocities and gradient of velocities. Because of the complexity of this theory - a complexity coming from the high number of the involved intrinsic and material parameters - the renaissance of the SGE theory takes place some decades later with the so-called simplified models where only two intrinsic parameters, one for microstructure and one for micro-inertia, are involved in the corresponding expressions of potential and kinetic energy densities, respectively. The sequence of those simplifications is explained in the works of Vardoulakis *et al.* (1996), Lazar *et al.* (2006), Askes and Aifantis (2011), Polizzoto (2017), Lurie *et al.* (2021) and Solyaev *et al.* (2022). The simplified SGE models have been extensively used to explain wave dispersion in materials with microstructure (e.g. Vardoulakis and Georgiadis, 1997; Georgiadis *et al.*, 2004; Papargyri-Beskou *et al.*, 2009; Vavva *et al.*, 2009; Berezovski *et al.*, 2011; Papacharalampopoulos *et al.*, 2011; Fafalis *et al.*, 2012; Gourgiotis *et al.*, 2013; Rosi *et al.*, 2018) and to solve related elastodynamic boundary value problems (e.g. Tsepoura *et al.*, 2002; Tsepoura and Polyzos, 2003; Papargyri-Beskou *et al.*, 2003; Polyzos *et al.*, 2005; Askes and Aifantis, 2011; Khakalo and Niiranen, 2018; Khakalo *et al.*, 2018; Hosseini and Niiranen, 2022), while different aspects of those models have been proposed and discussed by many investigators (e.g. Forest and Trinh, 2011; Javili *et al.*, 2013; Polizzoto, 2013a,b, 2014,2017; Broese *et al.*, 2016; Ojaghnezhada and Shodja, 2016; Lazar and Po, 2018).

Despite the intensive work on the SGE theories, one can say that still there is a lack of consensus among the scientists as to the answers to the two questions, first when the gradients of strains and velocities have to be taken into account and second what exactly represent the intrinsic parameters introduced by the strain and velocity gradient terms in the potential and kinetic energy. Many works appearing so far in the literature try to give convincing answers with experimental observations ( e.g. Dontsov *et al.*, 2013; Polyzos *et al.*, 2015; Lei *et al.*, 2016; Iliopoulos *et al.,* 2016, 2017), homogenization procedures between micro- and macroscales (e.g. Fish *et al.*, 2002; Kouznetsova *et al.*, 2002; Peerlings and Fleck, 2004; Auffray *et al.*, 2010; Triantafyllou and Giannakopoulos, 2013; Bacca *et al.*, 2013; Abdoul-Anziz, and Seppecher, 2018; Tan and Poh, 2018; Monchiet *et al.*, 2020) and continuation approaches of nonlocal mass-spring lattice models (e.g. Metrikine and Askes, 2002; Metrikine, 2006; Andrianov *et al.*, 2010; Fafalis et al., 2012; Polyzos and Fotiadis, 2012, 2019; De Domeniko et al. 2019; Gomez-Silva et al., 2020).

On the other hand, the concept of the NLE appears for the first time in the work of Kröner (1967) and afterwards is systematically introduced by Eringen and co-workers. The need for introducing a nonlocal elastic theory is adequately explained in Eringen (1999), Rogula (1982), Silling (2000), as well as in Beban and McCoy (1970) and Ostoja-Starzewski (2008) where statistical models applied to random non-homogeneous materials can be replaced by nonlocal ones valid for a single solid. Because of the integrodifferential form of the constitutive equations and the equation of motion of the NLE, its application to material and structural dynamics was a difficult task. To overcome that obstacle, Eringen (1983) utilizing fundamental solution type kernels in the integrals of the NLE, proposed a differential form of his theory known as nonlocal differential form or stress gradient elastic theory. Although that stress gradient form of the original nonlocal theory suffers by some paradoxes (Challamel and Wang, 2008; Challamel et al., 2014; Fernández-Sáez et al., 2016; Koutsoumaris et al., 2017), its presence indicates that a strain integral model is associated with a stress gradient model and as it is stated by Borino and Polizzotto (2014), analogously, a strain gradient model can be associated to a stress integral model. This statement is the kernel of the present work, which proposes a simple alternative nonlocal theory that derives, in the framework of one-dimensional wave propagation, all the known strain gradient elastic theories as special cases. This is accomplished by considering a nonlocality not on stresses but on the expressions of potential and kinetic energy densities. The paper is organized as follows: in the next section a short review on the application of the classical, first strain gradient, second strain gradient and nonlocal elastic theories to 1D wave propagation is reported. Next, the new nonlocal elastic theory is illustrated and all the procedures that lead to different types of SGE theories appearing so far in the literature are addressed.

## 2. Wave propagation in classical, first strain gradient, second strain gradient and nonlocal continuum

Confining our analysis to one-dimensional deformation, in the present section, the propagation of a plane wave in classical, strain gradient, and nonlocal elastic materials is explained. The goal of this demonstration is twofold, first to point out the differences of the nonclassical elastic theories and classical elasticity and second to remind the dispersion relations provided by those theories for a longitudinal plane wave.

## 2.1 Classical elastic material

Fulfilling the three assumptions for a classical elastic continuum one reaches to the following three fundamental equations:

$$
\begin{aligned}
&e(x,t) = \partial_x u(x,t) &&\text{(i)}\\
&v(x,t) = \partial_t u(x,t) &&\text{(ii)}\\
&\sigma(x,t) = E e(x,t) &&\text{(iii)}
\end{aligned}
\qquad (1)
$$

where $u, e, \sigma$ stand for displacement, strain and Cauchy stress, respectively, along the $x$-axis defined at point $x$ at time $t, \partial_x, \partial_t$ indicates differentiation with respect to space and time, respectively, and $E$ is the Young modulus of elasticity.

In view of Eq. (1) the expressions for the potential and kinetic energy densities $P(x,t)$ and $T(x,t)$, respectively, read

$$
\begin{aligned}
&P(x,t) = \frac{1}{2} E \left[ e(x,t) \right]^2 &&\text{(i)}\\
&T(x,t) = \frac{1}{2} \rho \left[ \partial_t u(x,t) \right]^2 &&\text{(ii)}
\end{aligned}
\qquad (2)
$$

with $\rho$ being the mass density of the elastic material.

Applying Newton's law or Hamilton's principle and ignoring body forces, the following equation of motion is obtained

$$
\partial_x \sigma(x,t) = \rho \partial_{tt} u(x,t), \quad x \in R, t > 0
\qquad (3)
$$

which in terms of displacements is written as

$$E\partial_{xx}u(x,t) = \rho\partial_{tt}u(x,t), \quad x \in R, t > 0 \tag{4}$$

The propagation of a longitudinal plane wave $u(x,t) = Ue^{i(kx-\omega t)}$, according to equation (4), provides a dispersion relation of the form

$$c_p = \frac{\omega}{k} = \frac{E}{\rho} \tag{5}$$

where $c_p, k, \omega$ stand for the phase velocity, the wavenumber, and the frequency of the propagating wave, respectively.

Relation (5) reveals that all plane waves of different wavelengths propagate at the same velocity $c_p$, which means no dispersion occurs. It is apparent that the classical theory of elasticity is not able to support wave dispersion and this disadvantage is, among others, the motivation for the development of enhanced elastic theories, which most of them introduce intrinsic parameters to capture wave dispersion in conventional materials.

### 2.2 Simple first strain gradient elastic material

Following the simplifications of Mindlin's theory (Mindlin, 1964) described in Papargyri-Beskou *et al.* (2009), Askes and Aifantis (2011), Polyzos and Fotiadis (2012, 2019), Iliopoulos *et al.* (2016), Polizzotto (2017) and Khakalo and Niiranen (2018), the departure point of the simple first strain gradient elastodynamic theory is the assumption that the potential and kinetic energy densities defined at a point *x* at time *t* are functions not only of strains and velocities but also of their gradients. This enhancement modifies equations (2) as follows:

$$
\begin{aligned}
P(x,t) &= \frac{1}{2}E\left[e(x,t)\right]^2 + \frac{1}{2}Eg^2\left[\partial_x e(x,t)\right]^2 &\text{(i)} \\
T(x,t) &= \frac{1}{2}\rho\left[\partial_t u(x,t)\right]^2 + \frac{1}{2}\rho h^2\left[\partial_{xt} u(x,t)\right]^2 &\text{(ii)}
\end{aligned}
\tag{6}
$$

with $g$, $h$ being the two intrinsic length scale parameters (units of length) associated with microelastic and microinertia effects, respectively. Apparently, for $g=h=0$ the classical elastic material is recovered.

Exploiting the energy densities (6), the three fundamental equations (1) are replaced by the following ones

$$e(x,t) = \partial_x u(x,t) \qquad\qquad\qquad \text{(i)}$$
$$v(x,t) = \partial_t u(x,t) \qquad\qquad\qquad \text{(ii)}$$
$$\sigma(x,t) = \frac{\partial P(x,t)}{\partial e} = E e(x,t) \qquad\qquad \text{(iii)} \qquad\qquad (7)$$
$$\mu(x,t) = \frac{\partial P(x,t)}{\partial(\partial_x e)} = g^2 E \partial_x e(x,t) \qquad\qquad \text{(iv)}$$

where $\sigma(x,t)$ still represents Cauchy stresses and $\mu(x,t)$ stands for the so-called double stresses introduced by the not ignored strain gradients in the expression of the potential energy density. The application of Hamilton's principle here provides the following equation of motion

$$\partial_x \tau(x,t) = \rho\left(1 - h^2 \partial_{xx}\right)\partial_{tt} u(x,t), \quad x \in R, t > 0 \qquad\qquad (8)$$

which in terms of displacements is written as

$$E\left(1 - g^2 \partial_{xx}\right)\partial_{xx} u(x,t) = \rho\left(1 - h^2 \partial_{xx}\right)\partial_{tt} u(x,t), \quad x \in R, t > 0 \qquad\qquad (9)$$

The stress $\tau(x,t)$ in (8) represents the total stress, which in view of (7) and (9) admits the form

$$\tau(x,t) = \sigma(x,t) - g^2 \partial_x \mu(x,t) = \left(1 - g^2 \partial_{xx}\right)\sigma(x,t) \qquad\qquad (10)$$

Obviously, equation (10) violates the second assumption of classical elasticity by introducing in the definition of stresses the curvature of Cauchy stresses around the point x as well as the third constitutive assumption by defining the stresses at point $x$ not only through strains but also via the gradient of strains. The introduction of the curvature of Cauchy stresses in (10), manifests the weakly nonlocal nature of the first strain gradient elastic theory since the definition of stress curvature requires the knowledge of stresses at the neighborhood of point $x$.

The propagation of a longitudinal plane wave $u(x,t) = Ue^{i(kx-\omega t)}$, according to equation (9), leads to the following dispersion relation

$$\frac{V_p}{c_p} = \sqrt{\frac{1+g^2 k^2}{1+h^2 k^2}} \overset{\bar{k}=gk}{=} \sqrt{\frac{1+\bar{k}^2}{1+\dfrac{h^2}{g^2}\bar{k}^2}} \tag{11}$$

Indicatively for $g/h = 1.1$ and $g/h = 0.9$, the corresponding dispersion curves are demonstrated in Fig. 1.

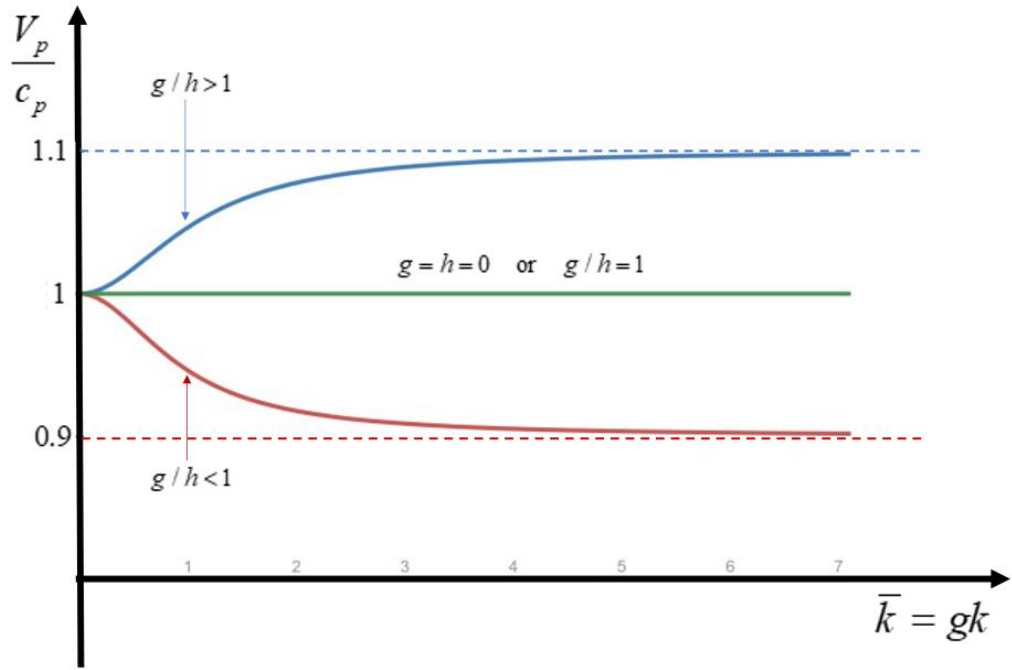

**Figure 1:** Wave dispersion in a first strain gradient elastic material. Phase velocity as a function of the dimensionless wave number for different values of the microelastic and microinertia internal length scale parameters $g$ and $h$, respectively.

Equation (11) reveals that longitudinal plane waves propagated in a first gradient elastic material undergo dispersion. That dispersion is entirely supported by the internal length scale parameters $g$ and $h$ which in case of $g=h=0$ any elastic pulse propagates undisturbed as in classical case. It should be noticed, however, that the classical elastic behavior is recovered not only in the case of $g=h=0$ but also when $g=h$, as it is apparent from the dispersion relation (11). This peculiarity will be discussed in the next section. As it is apparent from Fig. 1 and as it is discussed in Papargyri-Beskou *et al.* (2009), for $g>h$ the phase velocity of a longitudinal plane wave increases while the opposite trend is observed when $g<h$, a behavior which is also observed in the wave propagation in hardened and fresh concrete (Iliopoulos *et al.*, 2016). Taking the limit of equation (11) for $k \rightarrow \infty$, the upper or lower bound of the observed phase velocity is the ratio $g/h$. Finally, relation (11) is the only dispersion curve extracted by (9) indicating that the simple first strain gradient elastic theory does not support optical wave dispersion branches as the original strain gradient elastic theory of Mindlin does. A discussion on that can be found in Iliopoulos *et al.* (2016).

In case of $g=0$, the dispersion relation (11) obtains the form

$$\frac{V_p}{c_p} = \sqrt{\frac{1}{1+h^2k^2}} \qquad (12)$$

and its behavior is depicted in Fig.2.

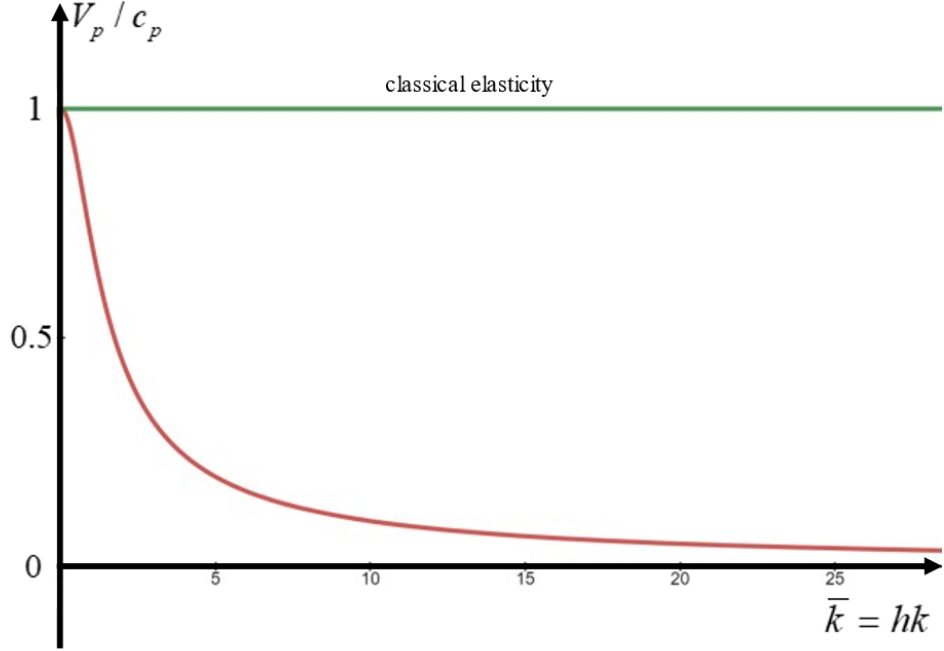

**Figure 2:** Wave dispersion of a plane wave according to dispersion relation (12).

Relation (12) indicates that only microinertia influences the propagation of a longitudinal plane wave and this happens only for small wave numbers. For high wave numbers the first gradient elastic material with $g$=0 does not support wave propagation.

### 2.3 *Simple second strain gradient elastic material*

Enhancing the expressions of potential and kinetic energy densities with second gradient strain and velocities terms, respectively, as it is explained in Polizzotto (2013b), Lazar *et al.* (2006) and Khakalo and Niiranen (2018), equations (6) are written in the form

$$P(x,t) = \frac{1}{2}E\left\{\left[e(x,t)\right]^2 + g^2\left[\partial_x e(x,t)\right]^2 + l^4\left[\partial_{xx}e(x,t)\right]^2 + 2Ce(x,t)\cdot\partial_{xx}e(x,t)\right\} \quad \text{(i)}$$

$$T(x,t) = \frac{1}{2}\rho\left[\partial_t u(x,t)\right]^2 + \frac{1}{2}\rho h^2\left[\partial_{xt}u(x,t)\right]^2 + \frac{1}{2}\rho d^4\left[\partial_{xxt}u(x,t)\right]^2 \quad \text{(ii)}$$

$$(13)$$

with *l*, *C*, *d* being new internal length scale parameters with $2C < g^2$.

Introducing (13) in Hamilton's variational relation, we obtain the following equation of motion

$$E\left[1-\left(g^2-2C\right)\partial_{xx}+l^4\partial_{xxxx}\right]\partial_{xx}u(x,t)=\rho\left[1-h^2\partial_{xx}+d^4\partial_{xxxx}\right]\partial_{tt}u(x,t),\quad x\in\square,t>0 \qquad (14)$$

The total stress $\tau(x,t)$ for the present case is written in terms of the classical Cauchy stresses as

$$\tau(x,t)=\sigma(x,t)-g^2\partial_{xx}\sigma(x,t)+l^4\partial_{xxxx}\sigma(x,t) \qquad (15)$$

As in the case of first strain gradient elasticity, the expression of total stresses (15) indicates that the second assumption of the classical elasticity is violated since the curvature of Cauchy stresses and its second space derivative are introduced in the final expression of stresses defined at the point $x$ and besides stresses are not defined via strains but also via gradient and higher order gradients of strains.

According to (14), the propagation of a longitudinal plane wave $u(x,t)=Ue^{i(kx-\omega t)}$ is affected by the following dispersion relation

$$\frac{V_p}{c_p}=\sqrt{\frac{1+\left(g^2-2C\right)k^2+l^4k^4}{1+h^2k^2+d^4k^4}}\Bigg|_{\bar{k}=gk}=\sqrt{\frac{1+\dfrac{g^2-2C}{g^2}\bar{k}^2+\dfrac{l^4}{g^4}\bar{k}^4}{1+\dfrac{h^2}{g^2}\bar{k}^2+\dfrac{d^4}{g^4}\bar{k}^4}} \qquad (16)$$

For $(g^2-2C)/g^2=0.9$ and $l^4/g^4=0.09$, Fig. 3 portrays the dispersion relation (16) for the pairs $g^2/h^2=1.211, d^4/g^4=0.0826$ and $g^2/h^2=0.826, d^4/g^4=0.1211$.

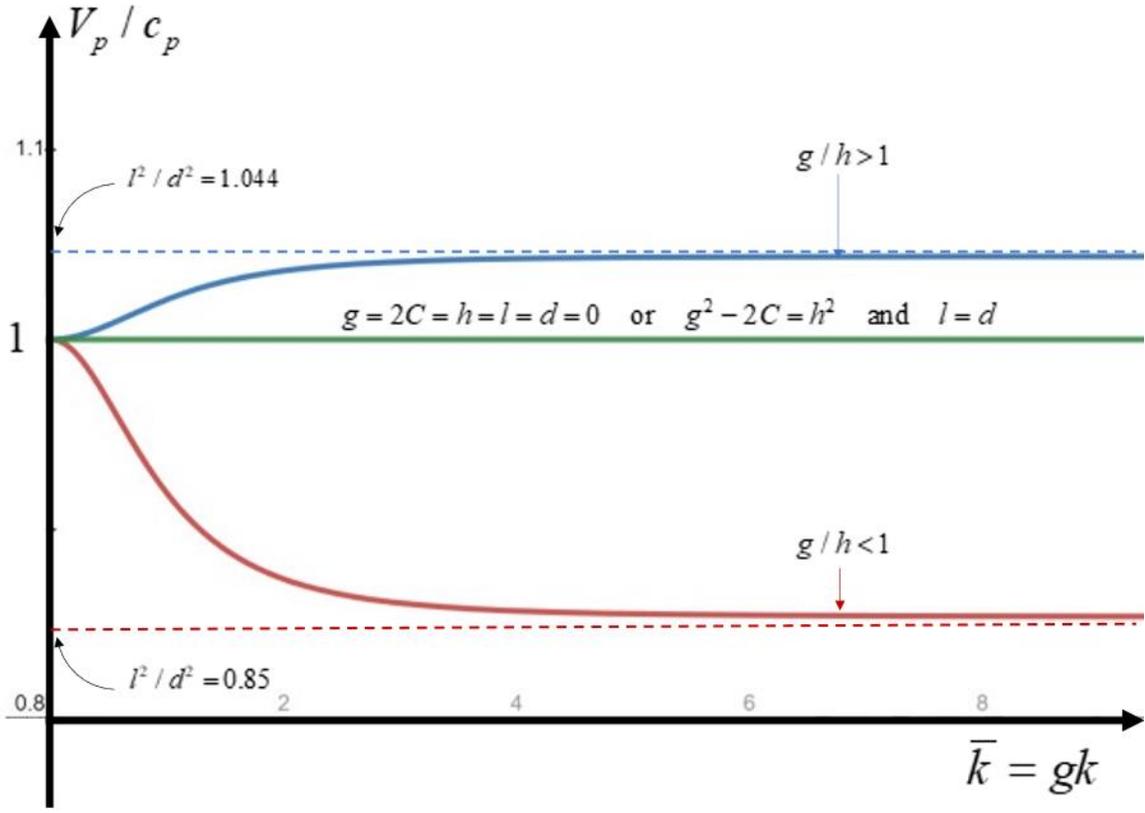

**Figure 3:** Wave dispersion in a second strain gradient elastic material. Phase velocity as a function of the dimensionless wave number for different values of the internal length scale parameters $g, C, l, h, d$ .

As in the case of first gradient elasticity, relation (16) addresses a wave dispersion governed by the internal length scale parameters *g, h, c, l, d.* Apparently the classical elasticity is recovered by either zeroing all those intrinsic parameters or considering $g^2 - 2c = h^2, l = d$ . For *g>h* the phase velocity of a longitudinal plane wave increases while the opposite occurs when *g<h*. The limit of equation (15) for $k \to \infty$ determines an upper and lower bound of the observed phase velocity equal to $l^2 / d^2$ .

## 2.4 Nonlocal elastic material

The nonlocal theory proposed by Eringen (1972, 1983), is a nonclassical elastic theory which violates the second fundamental assumption of classical elasticity by imposing a nonlocal nature on stresses. Thus, equations (1) are replaced by the following ones

$$
\begin{aligned}
&e(x,t) = \partial_x u(x,t) &&\text{(i)}\\
&v(x,t) = \partial_t u(x,t) &&\text{(ii)}\\
&\tau(x,t) = \int_{-\infty}^{\infty} K(|x-\xi|)\sigma(\xi,t)d\xi &&\text{(iii)}\\
&\sigma(x,t) = Ee(x,t) &&\text{(iv)}
\end{aligned}
\tag{17}
$$

and the equation of motion (3) is written in terms of the nonlocal stresses $\tau(x,t)$ as

$$
\partial_x \tau(x,t) = \rho \partial_{tt} u(x,t), \quad x \in \square, t > 0
\tag{18}
$$

The nonlocality is introduced via the integral $\int_{-\infty}^{\infty} K(|x-\xi|)\sigma(\xi,t)d\xi$, with a vanishing kernel $K(|x-\xi|)$ when $\xi$ is far from the source point $x$ and recovering the classical elastic form when $K(|x-\xi|)$ coincides with the Dirac delta function.

Denoting by $E$ the effective elastic modulus along the $x$ direction, then for a travelling plane wave $u(x,t) = Ue^{i(kx-\omega t)}$ the nonlocal integral of Eq. (17. iii) is written as

$$
\begin{aligned}
\int_{-\infty}^{\infty} K(|x-\xi|)\sigma(\xi,t)d\xi &= E\int_{-\infty}^{\infty} K(|x-\xi|)\frac{\partial}{\partial \xi}\left[Ue^{i(k\xi-\omega t)}\right]d\xi\\
&= E\int_{-\infty}^{\infty} K(|x-\xi|)(ik)Ue^{i(k\xi-\omega t)}d\xi = ikE\int_{-\infty}^{\infty} K(|x-\xi|)e^{ik(\xi-x)}Ue^{i(kx-\omega t)}d\xi\\
&= ikEUe^{i(kx-\omega t)}\int_{-\infty}^{\infty} K(\zeta)e^{ik\zeta}d\zeta = ikEUe^{i(kx-\omega t)}\sqrt{2\pi}\bar{K}(k)
\end{aligned}
\tag{19}
$$

where $\bar{K}(k)$ is the Fourier transform of the kernel function $K(\zeta)$.

In view of (19), the equation of motion (18) concludes to the following dispersion relation

$$\frac{V_p^2}{c_p^2} = \sqrt{2\pi}\,\bar{K}(k) \tag{20}$$

Different forms of the kernel function $K(\zeta)$ can be utilized for the nonlocal representation of stresses. The most widely used is that of Gaussian distribution (Eringen, 1983; Abdollahi and Boroomand, 2014), i.e.

$$K(\zeta) = \frac{1}{p\sqrt{\pi}} e^{-\frac{\zeta^2}{p^2}} \tag{21}$$

with $p$ controlling the width of the Gaussian bell.

The Fourier transform of (21) gives

$$\bar{K}(k) = \frac{1}{\sqrt{2\pi}} e^{-\frac{p^2 k^2}{4}} \tag{22}$$

and the dispersion relation (20) obtains the form

$$\frac{V_p^2}{c_p^2} = e^{-\frac{p^2 k^2}{4}} \tag{23}$$

The Taylor expansion of the exponential function of (23) and the assumption $a^2 k^2 \ll 1$ leads to a new dispersion relation of the form:

$$\frac{V_p^2}{c_p^2} = e^{-\frac{p^2 k^2}{4}} = 1 - \frac{p^2 k^2}{4} + O(k^4) \overset{p^2 k^2 \ll 1}{=} \frac{1}{1 + \frac{p^2 k^2}{4}} \tag{24}$$

Obviously, for $p^2 / 4 = h^2$ the above dispersion relation is the same with that of the first strain gradient elasticity (12). This similarity is the impetus for the new nonlocal theory presented in the next section.

Both dispersion relations, given by Eqs (23) and (24) are plotted in Fig. 4. Obviously, for $p$=0 the classical elastic behavior is obtained.

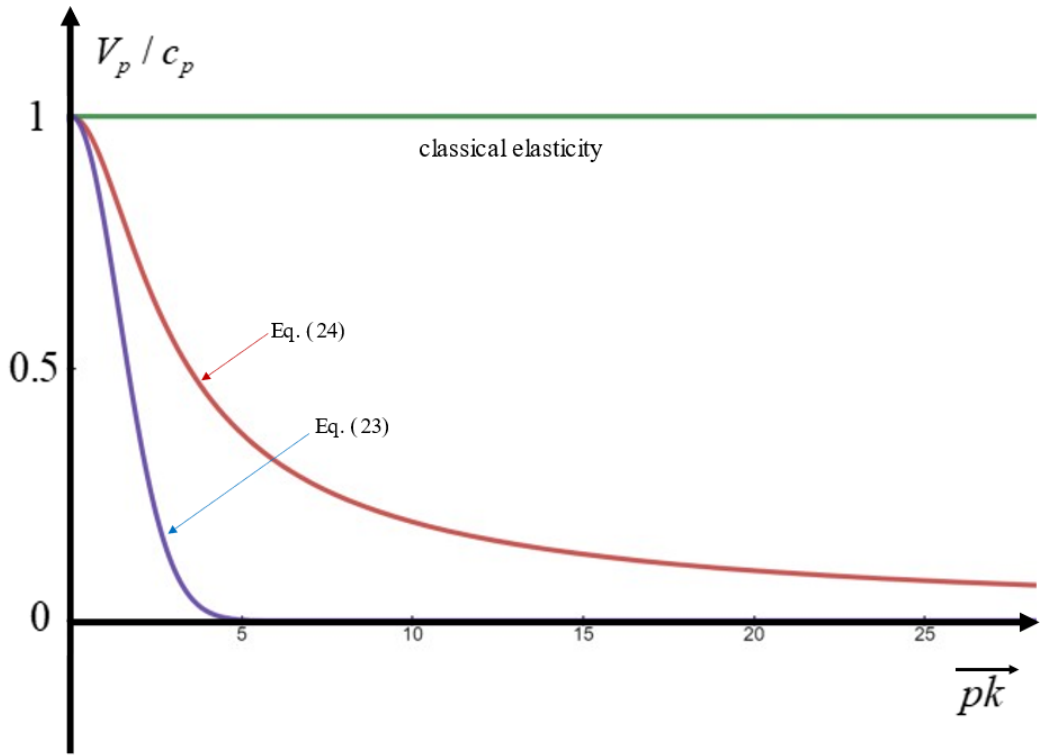

**Figure 4:** Wave dispersion in nonlocal elastic material. Phase velocity as a function of the wave number $k$ according to dispersion relations (22) and (23)

### 3. The nonlocal nature of the strain gradient elastic theories

This section is the kernel of the present work, and its goal is to propose an alternative nonlocal theory which gives all the known strain gradient elastic theories as special cases. Furthermore, provides an estimation of the internal length scale parameters used in those theories.

This specific nonlocality is expressed as follows:

$$\hat{P}(x,t) = \int\limits_{-a}^{a} K\left(\frac{\xi}{a}\right) P(x+\xi,t) d\xi$$

(25)

$$P(x,t) = \frac{1}{2} E e^2(x,t)$$

where $K\left(\dfrac{\xi}{a}\right)$ is a kernel possessing the Dirac-type property, i.e.

$$\int\limits_{-a}^{a} K\left(\frac{\xi}{a}\right) d\xi = 1$$

(26)

and the elastic modulus $E$ has the sam meaning as in the case of the classical nonlocal elastic theory.

The point $x$ is centered in the cell [-$a$, $a$] and the size $2a$ defines the nonlocal domain that influences the potential energy density at point $x$. In the classical nonlocal elastic theory, illustrated in the previous section, this nonlocal domain occupies all the available space of the considered continuum. Here, as in the case of peridynamics (Silling, 2000), there is a cutoff of the classical nonlocal space called horizon $a$. That nonlocal area could be a representative volume element (RVE) of a nonhomogeneous material, the unit cell of a periodic structure, or clusters of RVEs and unit cells since, as is pointed out in Forest (1998), neighboring RVEs or cells oppose resistance to the motion of the cell that contains the point $x$.

Following the suggestions of Eringen (1983) for 1D problems, in the present work the kernel $K\left(\dfrac{\xi}{a}\right)$ has been chosen to be equal to:

$$K\left(\frac{\xi}{a}\right) = \frac{1}{1.49364a} e^{-\frac{\xi^2}{a^2}}$$

(27)

so that the property (26) to be satisfied.

The kernel (27) is an even function of $\xi$ and subsequently the following relations hold true

$$\int_{-a}^{a} \xi K\left(\frac{\xi}{a}\right) d\xi = \int_{-a}^{a} \xi^3 K\left(\frac{\xi}{a}\right) d\xi = 0$$

$$\int_{-a}^{a} \xi^2 K\left(\frac{\xi}{a}\right) d\xi = \frac{a^2}{1.49364} \frac{e\sqrt{\pi}\, erf(1) - 2}{2e} = 0.2537 a^2$$

$$\int_{-a}^{a} \xi^4 K\left(\frac{\xi}{a}\right) d\xi = \frac{a^4}{1.49364} \frac{3e\sqrt{\pi}\, erf(1) - 10}{4e} = 0.13426 a^4$$

(28)

with $erf(x)$ being the Error function.

Almost all the nonlocal theories appearing so far in the literature do not consider any nonlocality in the kinetic energy density. However, Dayal (2017) in his work on peridynamics proposed a spatial nonlocality in the kinetic energy density to balance the spatial nonlocality of the potential energy density. In the same spirit, we introduce a spatial nonlocality in kinetic energy density in the cell [-b, b] as follows

$$\hat{T}(x,t) = \frac{1}{2b}\left[\int_{-b}^{b} T(x+\xi,t) d\xi\right]$$

$$T(x,t) = \frac{1}{2}\rho\left[\partial_t u(x,t)\right]^2$$

(29)

Equation (25) implies that the potential energy density defined at point $x$ is considered as a Gaussian distribution of all the corresponding densities defined in the space near to the point $x$, $A : x \in [-a, a]$, while equation (29) assumes the kinetic energy density as the average of the kinetic energy densities appearing in the domain $B : x \in [-b, b]$. In other words, the nonlocality in potential energy density is due to the action of long-range dynamic energy packets, like the long-range forces, while the nonlocality of the kinetic energy density comes from the contribution of the momentum packages lying inside to horizon $b$. Obviously, the two cells $[-a, a], [-b, b]$ are in general different from each other.

Next, we consider the Taylor series expansions for displacements and velocities around the point $x$, i.e.

$$u(x+\xi,t) = u(x,t) + \xi\partial_x u(x,t) + \frac{1}{2}\xi^2\partial_{xx}u(x,t) + \frac{1}{6}\xi^3\partial_{xxx}u(x,t) + \frac{1}{24}\xi^4\partial_{xxxx}u(x,t) + \mathrm{O}(\xi^5)$$

$$e(x+\xi,t) = \partial_x u(x,t) + \xi\partial_{xx}u(x,t) + \frac{1}{2}\xi^2\partial_{xxx}u(x,t) + \frac{1}{6}\xi^3\partial_{xxxx}u(x,t) + \mathrm{O}(\xi^4)$$

$$\upsilon(x+\xi,t) = \partial_t u(x,t) + \xi\partial_{xt}u(x,t) + \frac{1}{2}\xi^2\partial_{xxt}u(x,t) + \frac{1}{6}\xi^3\partial_{xxxt}u(x,t) + \mathrm{O}(\xi^4)$$

or for the sake of brevity

$$e(x+\xi,t) = u' + \xi u'' + \frac{1}{2}\xi^2 u''' + \frac{1}{6}\xi^3 u'''' + \mathrm{O}(\xi^3) \tag{30}$$

$$\upsilon(x+\xi,t) = \dot{u} + \xi\dot{u}' + \frac{1}{2}\xi^2\dot{u}'' + \frac{1}{6}\xi^3\dot{u}''' + \mathrm{O}(\xi^3) \tag{31}$$

and examine the following nonlocal models:

**Model I:** We assume that strains and velocities are constant in the nonlocal spaces $A: x \in [-a,a]$ and $B: x \in [-b,b]$, respectively. Then, keeping the constant terms of the expansions (30) and (31), the nonlocal potential energy expression (25) obtains the form

$$\begin{aligned}
\hat{P}(x,t) &= \int_{-a}^{a} K\left(\frac{\xi}{a}\right) P(x+\xi,t)d\xi = \int_{-a}^{a} K\left(\frac{\xi}{a}\right)\frac{1}{2}Ee^2(x+\xi,t)d\xi \\
&= \int_{-a}^{a} K\left(\frac{\xi}{a}\right)\frac{1}{2}E[u']^2\, d\xi \\
&= \frac{E}{2}[u']^2\left[\int_{-a}^{a} K\left(\frac{\xi}{a}\right)d\xi\right] \overset{Eq.(27)}{=} \frac{1}{2}E[u']^2
\end{aligned} \tag{32}$$

Similarly, for the expression of the kinetic energy (29) we obtain

$$\begin{aligned}
\hat{T}(x,t) &= \frac{1}{2b}\left[\int_{-b}^{b}\frac{1}{2}\rho[\upsilon(x+\xi,t)]^2 d\xi\right] = \frac{1}{2b}\left[\int_{-b}^{b}\frac{\rho}{2}[\dot{u}]^2 d\xi\right] \\
&= \frac{1}{2b}\frac{\rho}{2}[\dot{u}]^2[\xi]_{-b}^{b} = \frac{1}{2}\rho\dot{u}^2
\end{aligned} \tag{33}$$

The "action" function for the Model I has the form

$$L = \hat{T}(x,t) - \hat{P}(x,t) = \frac{1}{2}\rho \dot{u}^2 - \frac{1}{2}E(u')^2 \tag{34}$$

and the Hamilton's principle provides the following equation of motion

$$-\partial_t \left[\frac{\partial L}{\partial \dot{u}}\right] - \partial_x \left[\frac{\partial L}{\partial u'}\right] = 0 \Rightarrow Eu'' - \rho \ddot{u} = 0 \tag{35}$$

The equation of motion (35) is the same with that of (4), which means that Model I is nothing else than the classical theory of elasticity. The main conclusion here is that when strains and velocities vary constantly in the corresponding nonlocal spaces, then the nonlocality of potential and kinetic energy densities is not capable to disturb the local assumptions of the classical elastic theory.

**Model II:** Strains remain constant in the space $A: x \in [-a,a]$, while velocities vary linearly in the nonlocal domain $B: x \in [-b,b]$. In this case the potential energy density is provided again by Eq. (32) while the kinetic energy density obtains the form

$$
\begin{aligned}
\hat{T}(x,t) &= \frac{1}{2b}\left[\int_{-b}^{b}\frac{1}{2}\rho[\upsilon(x+\xi,t)]^2 d\xi\right] = \frac{1}{2b}\left[\int_{-b}^{b}\frac{1}{2}\rho[\dot{u}+\xi\dot{u}']^2 d\xi\right] \\
&= \frac{1}{2b}\left[\frac{\rho}{2}\int_{-b}^{b}\left\{[\dot{u}]^2 + 2\xi\dot{u}\cdot\dot{u}' + \xi^2[\dot{u}']^2\right\}d\xi\right] \\
&= \frac{1}{2b}\frac{\rho}{2}\dot{u}^2 2b + \frac{1}{2b}\frac{\rho}{2}[\dot{u}']^2\frac{2b^3}{3} = \frac{1}{2}\rho\dot{u}^2 + \frac{1}{2}\rho\left(\frac{b^2}{3}\right)[\dot{u}']^2
\end{aligned} \tag{36}
$$

In view of (32) and (36) the Hamilton's principle leads to the following equation of motion

$$-\partial_t\left[\frac{\partial L}{\partial\dot{u}}\right]+\partial_{tx}\left[\frac{\partial L}{\partial\dot{u}'}\right]-\partial_x\left[\frac{\partial L}{\partial u'}\right]=0\Rightarrow Eu''-\rho\left(\ddot{u}-\frac{b^2}{3}\ddot{u}''\right)=0 \qquad (37)$$

The dispersion relation corresponding to the Eq. (37) is

$$\frac{V_p^2}{c_p^2}=\frac{1}{1+\dfrac{b^2k^2}{3}} \qquad (38)$$

Eq. (37) and its dispersion relation (38) reveals that the Model II represents the first strain gradient elastic theory for $g$=0 and $h^2=b^2/3$. In addition to, the relation (38) coincides to the low frequency nonlocal dispersion relation (24) when $b^2=3p^2/4$. The importance of (38) is reflected by the fact that the intrinsic parameter $b^2/3$ is not just a normalization constant in the expression of kinetic energy density, as the $h^2$ in Eq. (6ii), but gives a measure of the nonlocal elastic area that influences the dynamic behavior of the material at point $x$. Also, as pointed out in Tan and Poh (2018), the nonuniformity in velocity within the nonlocal cell $A:x\in[-a,a]$ is perceived at the macro-level as a microinertia effect. It should be mentioned here that dispersion relation (38) is a result of the linear variation of displacements and velocities in the corresponding nonlocal spaces. Thus, it is not correct to say that for $b$=0 the classical elastic case is recovered. The classical case corresponds only to Model I where strains and velocities vary constantly in the corresponding nonlocal cells. The same statement is valid for all the models demonstrated in the present section.

**Model III:** Both strains and velocities vary linearly in the spaces $A:x\in[-a,a]$ and $B:x\in[-b,b]$, respectively. For this case, the kinetic energy density remains the same as in Eq. (36) and the nonlocal potential energy density is written as follows:

$$\hat{P}(x,t) = \int\limits_{-a}^{a} K\left(\frac{\xi}{a}\right) P(x+\xi,t)d\xi = \int\limits_{-a}^{a} K\left(\frac{\xi}{a}\right)\frac{1}{2}E\left[e(x+\xi,t)\right]^2 d\xi$$

$$= \int\limits_{-a}^{a} K\left(\frac{\xi}{a}\right)\frac{1}{2}E\left[u'+\xi u''\right]^2 d\xi$$

$$= \int\limits_{-a}^{a} K\left(\frac{\xi}{a}\right)\frac{1}{2}E\left\{\left[u'\right]^2 + 2\xi u'\cdot u'' + \xi^2\left[u''\right]^2\right\} d\xi$$

$$= \frac{1}{2}E\left[u'\right]^2\int\limits_{-a}^{a} K\left(\frac{\xi}{a}\right)d\xi + \frac{1}{2}E\left[u''\right]^2\int\limits_{-a}^{a} \xi^2 K\left(\frac{\xi}{a}\right)d\xi$$

$$\overset{Eq.(27)}{=} \frac{1}{2}E\left[u'\right]^2 + 0.2537 a^2\frac{1}{2}E\left[u''\right]^2 \square \frac{1}{2}E\left[\partial_x u\right]^2 + \frac{1}{2}E\frac{a^2}{4}\left[\partial_{xx}u\right]^2 \tag{39}$$

Applying the Hamilton's principle for (36) and (39), we obtain the following equation of motion

$$-\partial_t\left[\frac{\partial L}{\partial \dot{u}}\right] + \partial_{tx}\left[\frac{\partial L}{\partial \dot{u}'}\right] - \partial_x\left[\frac{\partial L}{\partial u'}\right] + \partial_{xx}\left[\frac{\partial L}{\partial u''}\right] = 0 \Rightarrow$$

$$E\left(u'' - \frac{a^2}{4}u''''\right) - \rho\left(\ddot{u} - \frac{b^2}{3}\ddot{u}''\right) = 0 \tag{40}$$

accompanied by the dispersion relation

$$\frac{V_p}{c_p} = \sqrt{\frac{1+\dfrac{a^2}{4}k^2}{1+\dfrac{b^2}{3}k^2}} \tag{41}$$

It is evident that Eq. (40) represents the equation of motion of the first strain gradient elastic theory given by Eq. (9). Besides, Eq. (40) provides an estimation of the intrinsic parameters $g$ and $h$ appearing in Eq. (9) with respect to the size of the nonlocal spaces used for the definition of the potential and kinetic energy densities at point $x$, i.e., $g^2 = a^2/4$ and $h^2 = b^2/3$. As in the case of Lagrange action function, it seems that the dynamic behavior of a material with microstructural effects is a result of the competition of two distinguished underline mechanisms, the micro-elasticity and micro-inertia expressed via the nonlocal cells $2a$ and $2b$, respectively.

Recalling the dispersion relation (11), it is apparent that when $a > 2b/\sqrt{3}$ the phase velocity of a longitudinal plane wave increases as the wave number increases while the opposite happens when $a < 2b/\sqrt{3}$. In case where $a = 2b/\sqrt{3}$, it is impressive that the material behaves like a classical elastic material. This classical behavior is likely the result of the cancellation of microstructural effect by the microinertia one. In addition, the nonlocal formation of Model III gives an answer on when the gradient of strains and the gradient of velocities have to be taken into account in the expression of potential and kinetic energy densities, respectively. The derivation of Eq. (40) is also accomplished in Polyzos and Fotiadis (2012) via a continuation approach of a lattice model containing masses and springs with distributed mass. In that lattice model the nonlocal parameter $a$ corresponds to the size of the lattice unit cell $L$ and $b = L\sqrt{\rho'/\rho}$ with $\rho', \rho$ being the density of microstructure and macrostructure, respectively. However, the difficulty of defining the correspondence between lattice unit cell and material microstructure renders the present Model III more practical than the corresponding mass-springs lattice models. From the above analysis, it is obvious that for quadratic variation of displacements in the space $A : x \in [-a, a]$ and constant velocities in $B : x \in [-b, b]$, the dispersion relation (40) is reduced to

$$\frac{V_p}{c_p} = \sqrt{1 + \frac{a^2}{4}k^2} = \sqrt{1 + 0.25\overline{k}^2} \qquad (42)$$

which shows an unrealistic wave behavior with continuously increasing velocity as the wave number increases. According to Model III, this means that it is not compatible to have one linear variation of strains and constant velocities in the nonlocal areas of point $x$.

**Model IV:** We consider linear variation of strains in the nonlocal spaces $A : x \in [-a, a]$ and linear variation of velocities in the space and time intervals, $B : x \in [-b, b]$ and $C : t \in [-c, c]$, respectively, i.e.,

$$e(x+\xi,t) = u' + \xi u''$$ (43)

$$\dot{u}(x+\xi,t+\tau) = \dot{u} + \xi \dot{u}' + \tau \ddot{u}$$ (44)

Then, the nonlocal potential energy density remains the same as in the Model III, i.e.,

$$\hat{P}(x,t) = \frac{1}{2} E [\partial_x u]^2 + \frac{1}{2} E \frac{a^2}{4} [\partial_{xx} u]^2$$ (45)

while the nonlocal kinetic energy density is formed as

$$\begin{aligned}
\hat{T} &= \frac{1}{2b} \int_{-b}^{b} \frac{1}{2c} \int_{-c}^{c} \frac{1}{2} \rho [\upsilon(x+\xi,t+\tau)]^2 \, d\tau \, d\xi \\
&= \frac{1}{2b} \int_{-b}^{b} \frac{1}{2c} \int_{-c}^{c} \frac{1}{2} \rho [\dot{u} + \xi \dot{u}' + \tau \ddot{u}]^2 \, d\tau \, d\xi \\
&= \frac{1}{4bc} \int_{-b}^{b} \int_{-c}^{c} \frac{1}{2} \rho \left[ \dot{u}^2 + \xi^2 (\dot{u}')^2 + \tau^2 \ddot{u}^2 \right] d\tau \, d\xi \\
&= \frac{1}{2} \rho \dot{u}^2 + \frac{1}{2} \rho \frac{b^2}{3} [\dot{u}']^2 + \frac{1}{2} \rho \frac{c^2}{3} \ddot{u}^2
\end{aligned}$$ (46)

with c having units of time.

The Hamilton's principle states that

$$-\partial_t \left[ \frac{\partial L}{\partial \dot{u}} \right] + \partial_{tx} \left[ \frac{\partial L}{\partial \dot{u}'} \right] + \partial_{tt} \left[ \frac{\partial L}{\partial \ddot{u}} \right] - \partial_x \left[ \frac{\partial L}{\partial u'} \right] + \partial_{xx} \left[ \frac{\partial L}{\partial u''} \right] = 0$$ (47)

and leads to the following equation of motion

$$E \left[ u'' - \frac{a^2}{4} u'''' \right] - \rho \left( \ddot{u} - \frac{b^2}{3} \ddot{u}'' - \frac{c^2}{3} \ddddot{u} \right) = 0$$ (48)

The dispersion relation that corresponds to (48) and concerns the propagation of a longitudinal plane wave $u(x,t) = U e^{i(kx - \omega t)}$ reads

$$E\left(-k^2 - \frac{a^2}{4}k^4\right) = \rho\left(-\omega^2 - \frac{b^2}{3}\omega^2 k^2 - \frac{c^2}{3}\omega^4\right) \Rightarrow$$

$$\left[\frac{1}{3}\left(\frac{\hat{c}}{a}\right)^2\right]\Omega^4 + \left[1 + \frac{1}{3}\left(\frac{b}{a}\right)^2 \bar{k}^2\right]\Omega^2 - \bar{k}^2\left(1 + \frac{1}{4}\bar{k}^2\right) = 0 \Rightarrow \qquad (49)$$

$$\left[\frac{1}{3}\left(\frac{\hat{c}}{a}\right)^2 \bar{k}^2\right]\frac{V_p^4}{c_p^4} + \left[1 + \frac{1}{3}\left(\frac{b}{a}\right)^2 \bar{k}^2\right]\frac{V_p^2}{c_p^2} - \left(1 + \frac{1}{4}\bar{k}^2\right) = 0$$

and provides in terms of frequency vs wavelength and phase velocity vs wavelength the following real roots, respectively

$$\Omega = \sqrt{\frac{-\left[1 + 0.333\left(\frac{b}{a}\right)^2 \bar{k}^2\right] + \sqrt{\left[1 + 0.333\left(\frac{b}{a}\right)^2 \bar{k}^2\right]^2 + 1.333\left(\frac{\hat{c}}{a}\right)^2 \bar{k}^2\left(1 + 0.25\bar{k}^2\right)}}{2 \times 0.333\left(\frac{\hat{c}}{a}\right)^2}} \qquad (50)$$

$$\frac{V_p}{c_p} = \sqrt{\frac{-\left[1 + 0.333\left(\frac{b}{a}\right)^2 \bar{k}^2\right] + \sqrt{\left[1 + 0.333\left(\frac{b}{a}\right)^2 \bar{k}^2\right]^2 + 1.333\left(\frac{\hat{c}}{a}\right)^2 \bar{k}^2\left(1 + 0.25\bar{k}^2\right)}}{2 \times 0.333\left(\frac{\hat{c}}{a}\right)^2 \bar{k}^2}} \qquad (51)$$

where $\Omega = \frac{\omega^2 a^2}{c_p^2}$, $\bar{k} = ak$, and $\hat{c} = cc_p$.

The plots of (50) and (51) for different pairs of $\left(\frac{b}{a}\right)^2$ and $\left(\frac{\hat{c}}{a}\right)^2$ is provided in Fig 5.

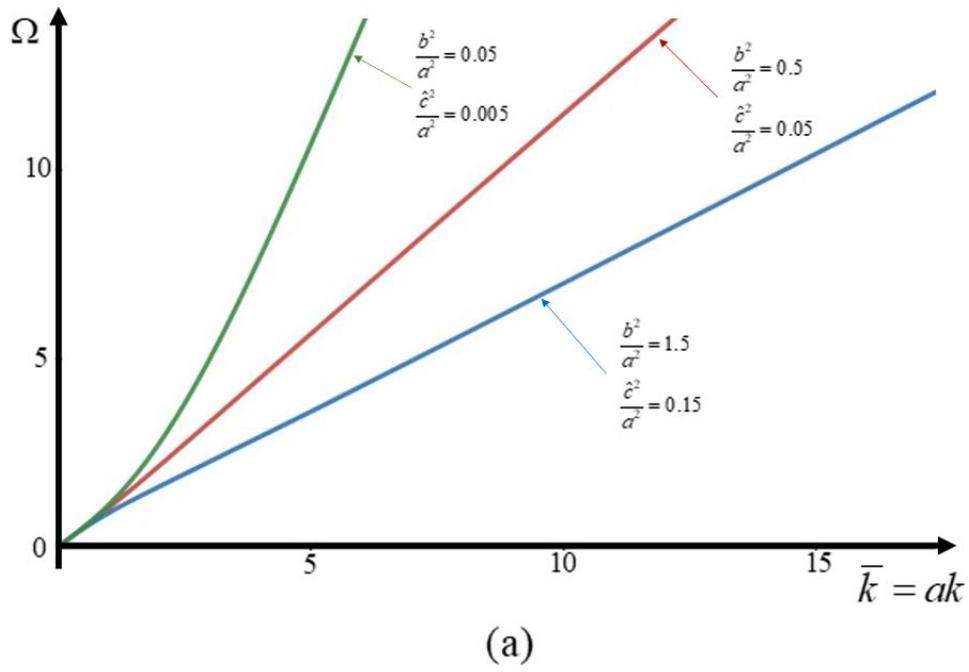

(a)

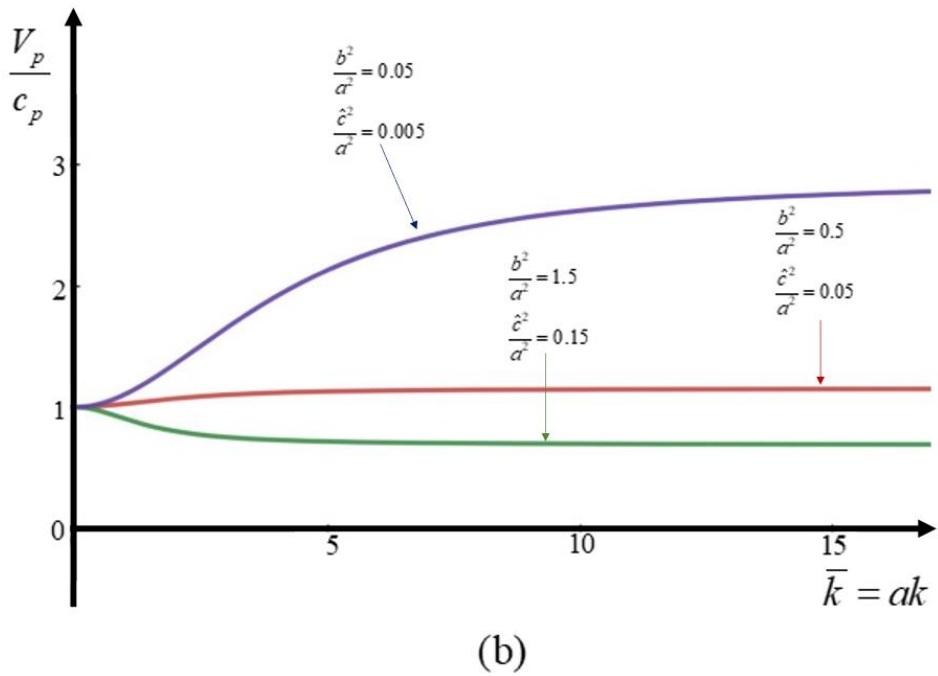

(b)

**Figure 5:** (a) The dispersion relation (50) and (b) the dispersion relation (51), for different values of $\left(b/a\right)^2$ and $\left(\hat{c}/a\right)^2$.

Metrikine (2006) invoking Einstein's causality principle, claimed that the partial differential equation that describes the behavior of propagating waves in materials with microstructural effects must contain derivatives of the same order in space and time. To this goal, he added in the equation of motion of Model III a fourth-order derivative of displacements with respect to time without providing systematic proof of that. Later, Pichugin et al. (2008) applying second partial derivatives of time in asymptotic expansions of displacements, derived heuristically an equation of motion that contains a fourth-order time derivative of displacements. That model was employed by Dontsov et al. (2013) to evaluate through analytical treatment and experimental measurements the corresponding internal length scale parameters for a longitudinal wave propagating in a material with periodic microstructure. The present Model IV introduces, in addition to space nonlocality, a memory mechanism in the definition of the nonlocal kinetic energy density thus offering a complete theoretical derivation of the causal equation of motion used in the above-mentioned works. However, the terms containing the fourth-order time derivatives in the above-mentioned works and in Eq. (48) have opposite signs. This difference has an important effect on the corresponding dispersion curves, i.e., the model of Metrikine (2006) provides a dispersion curve with two branches while Eq. (48) addresses a stable solution with a single dispersion curve like those depicted in Fig. 5. The provided dispersion relations of the present model are like those provided by Model III with the difference that

$$\lim_{k \to \infty} \left( \frac{V_p}{c_p} \right)_{\text{Model III}} = \frac{\sqrt{3}}{2} \frac{a}{b}$$

$$\lim_{k \to \infty} \left( \frac{V_p}{c_p} \right)_{\text{Model IV}} = \frac{\sqrt{2}}{2} \frac{b}{\hat{c}} \sqrt{-1 + \sqrt{1 + 3 \left( \frac{\hat{c}}{b} \right)^2 \left( \frac{a}{b} \right)^2}}$$

**Model V:** Strains remain linear, and velocities vary quadratically in the spaces $A : x \in [-a, a]$ and $B : x \in [-b, b]$, respectively. The potential energy density is provided by Eq. (45), while the kinetic energy density admits the form:

$$\hat{T} = \frac{1}{2b}\left[\int_{-b}^{b}\frac{1}{2}\rho[\upsilon(x+\xi,t)]^2\,d\xi\right] = \frac{1}{2b}\left[\int_{-b}^{b}\frac{1}{2}\rho\left[\dot{u}+\xi\dot{u}'+\frac{1}{2}\xi^2\dot{u}''\right]^2 d\xi\right]$$

$$= \frac{1}{2b}\left[\frac{\rho}{2}\int_{-b}^{b}\left\{\dot{u}^2+\xi^2\dot{u}\cdot\dot{u}''+\xi^2[\dot{u}']^2+\frac{\xi^4}{4}[\dot{u}'']^2\right\}d\xi\right]$$

$$= \frac{1}{2b}\frac{\rho}{2}\dot{u}^2 2b+\frac{1}{2b}\frac{\rho}{2}\left\{\dot{u}\cdot\dot{u}''+[\dot{u}']^2\right\}\frac{2b^3}{3}+\frac{1}{2b}\frac{\rho}{2}[\dot{u}'']^2\frac{2b^5}{20}$$

$$= \frac{1}{2}\rho\dot{u}^2+\frac{1}{2}\rho\left(\frac{b^2}{3}\right)\left\{\dot{u}\cdot\dot{u}''+[\dot{u}']^2\right\}+\frac{1}{2}\rho\left(\frac{b^4}{20}\right)[\dot{u}'']^2 \qquad (52)$$

Applying the Hamilton's principle for (45) and (52), we obtain the following equation of motion

$$-\partial_t\left[\frac{\partial L}{\partial\dot{u}}\right]+\partial_{tx}\left[\frac{\partial L}{\partial\dot{u}'}\right]-\partial_{txx}\left[\frac{\partial L}{\partial\dot{u}''}\right]-\partial_x\left[\frac{\partial L}{\partial u'}\right]+\partial_{xx}\left[\frac{\partial L}{\partial u''}\right]=0\Rightarrow$$

$$E\left(u''-\frac{a^2}{4}u''''\right)-\rho\left(\ddot{u}+\frac{b^4}{20}\ddot{u}''''\right)=0 \qquad (53)$$

which for a longitudinal plane wave provides the next dispersion relation

$$\frac{V_p}{c_p}=\sqrt{\frac{1+\dfrac{a^2}{4}k^2}{1+\dfrac{b^4}{20}k^4}}\;\overset{\bar{k}=ak}{=}\;\sqrt{\frac{1+0.25\bar{k}^2}{1+0.05\dfrac{b^4}{a^4}\bar{k}^4}} \qquad (54)$$

and for different values of the ratio $b^4/a^4$ is depicted in Fig. 6.

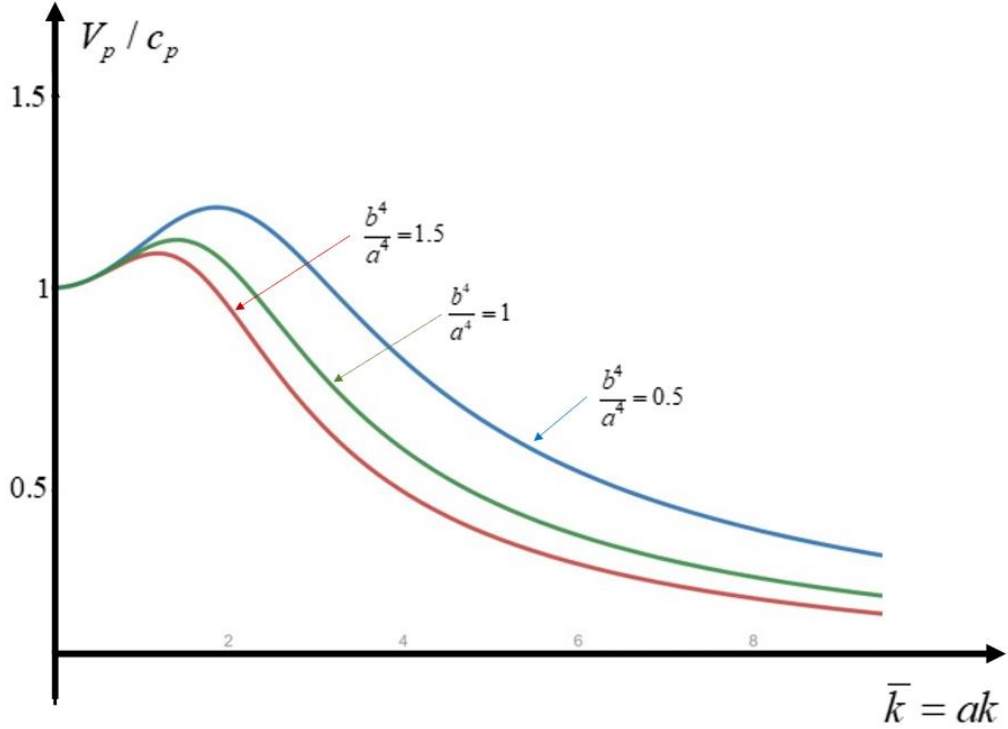

**Figure 6:** Wave dispersion in nonlocal Model V. Phase velocity as a function of the dimensionless wave number for different values of the ratio $b^4 / a^4$.

This model seems to be a degenerate case of the second strain gradient elastic theory since Eq. (54) can be taken by the Eq. (14) for $C = l = h = 0$. Although the present Model V and Model III employ two internal length scale parameters, it is apparent that in the present model the micro-inertia mechanism is always stronger than that of micro-elasticity. Because of that, it is not possible a non-dispersive wave propagation as it happens in Model III for specific values of the two intrinsic parameters. In the case where strains are constant and velocities vary quadratically in the corresponding nonlocal domains, the dispersion relation (54) is simplified to

$$\frac{V_p}{c_p} = \sqrt{\frac{1}{1 + \dfrac{b^4}{20} k^4}} \tag{55}$$

Equation (55) has been employed by De Domenico *et al.* (2019) to verify experimental measurements dealing with the propagation of longitudinal waves in bismuth.

**Model VI:** Strains and velocities have a quadratic variation in the spaces $A : x \in [-a, a]$ and $B : x \in [-b, b]$, respectively. In this case the kinetic energy density is given by Eq. (52), while the potential energy density is the following one:

$$
\begin{aligned}
\hat{P}(x,t) &= \int_{-a}^{a} K\left(\frac{\xi}{a}\right) P(x+\xi,t)\, d\xi = \int_{-a}^{a} K\left(\frac{\xi}{a}\right) \frac{1}{2} E\, [e(x+\xi,t)]^2\, d\xi \\
&= \int_{-a}^{a} K\left(\frac{\xi}{a}\right) \frac{1}{2} E\left[ u' + \xi u'' + \frac{1}{2}\xi^2 u''' \right]^2 d\xi \\
&= \int_{-a}^{a} K\left(\frac{\xi}{a}\right) \frac{1}{2} E\left\{ [u']^2 + \xi^2 [u'']^2 + \xi^2 u' \cdot u''' + \frac{\xi^4}{4} [u''']^2 \right\} d\xi \\
&\overset{Eq.(28)}{=} \frac{1}{2} E [u']^2 + 0.2537 a^2 \frac{1}{2} E\left\{ [u'']^2 + u' \cdot u''' \right\} + \frac{0.13426}{4} a^4 \frac{1}{2} E [u''']^2 \\
&\simeq \frac{1}{2} E [u']^2 + \frac{1}{2} E \frac{a^2}{4}\left\{ [u'']^2 + u' \cdot u''' \right\} + \frac{1}{2} E \frac{a^4}{30} [u''']^2
\end{aligned}
\tag{56}
$$

Applying the Hamilton's principle for (52) and (56), we obtain the following equation of motion

$$
\begin{aligned}
&-\partial_t\left[\frac{\partial L}{\partial \dot{u}}\right] + \partial_{tx}\left[\frac{\partial L}{\partial \dot{u}'}\right] - \partial_{txx}\left[\frac{\partial L}{\partial \dot{u}''}\right] - \partial_x\left[\frac{\partial L}{\partial u'}\right] + \partial_{xx}\left[\frac{\partial L}{\partial u''}\right] - \partial_{xxx}\left[\frac{\partial L}{\partial u'''}\right] = 0 \Rightarrow \\
&E\left( u'' + \frac{a^4}{30} u'''''' \right) - \rho\left( \ddot{u} + \frac{b^4}{20} \ddot{u}'''' \right) = 0
\end{aligned}
\tag{57}
$$

and the following dispersion relation

$$
\frac{V_p}{c_p} = \sqrt{\frac{1 + \dfrac{a^4}{30} k^4}{1 + \dfrac{b^4}{20} k^4}} \overset{\bar{k}=ak}{=} \sqrt{\frac{1 + 0.033\, \bar{k}^4}{1 + 0.05\, \dfrac{b^4}{a^4}\, \bar{k}^4}}
\tag{58}
$$

which for $b^4 / a^4 = 1.5, 1.0, 1.5$ is represented in Fig. 7.

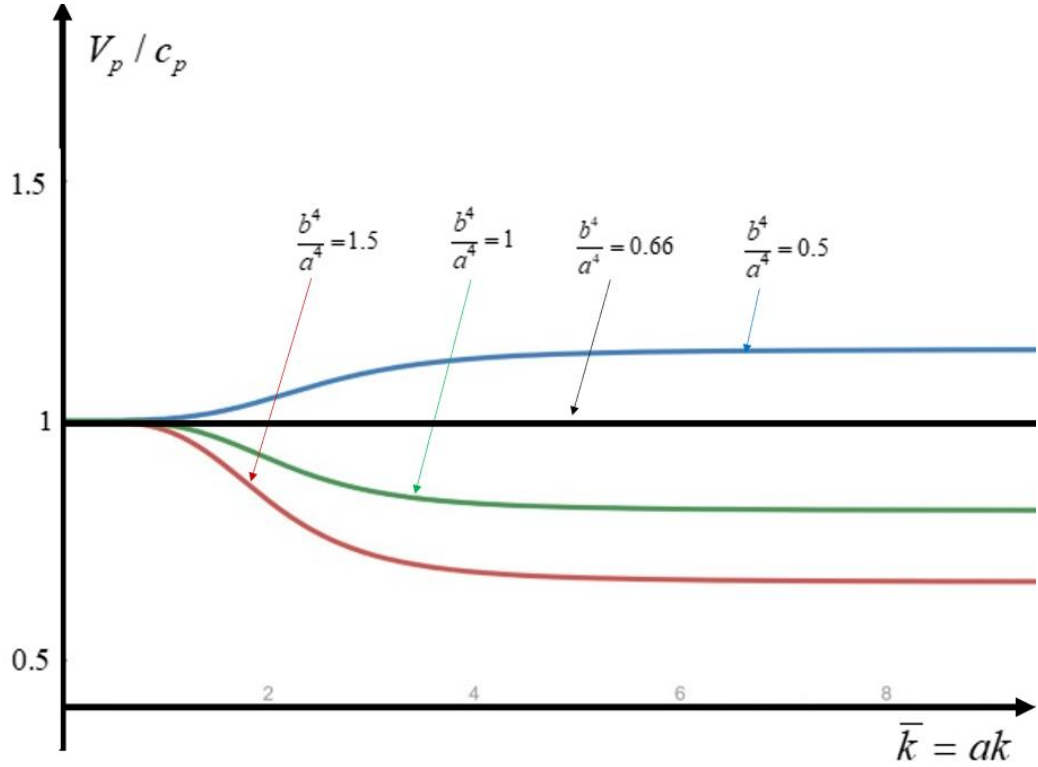

**Figure 7:** Wave dispersion relation (44). Phase velocity as a function of the dimensionless wave number for different values of the ratio $b^4 / a^4$.

The model VI is a higher order strain gradient model with two internal length scale parameters depended on the horizons *a*, *b*. As it is evident from Eq. (58) and Fig. 7, for large wavenumbers the longitudinal phase velocity appears a plateau approaching the limit $\sqrt{2/3}\, a^2 / b^2$. Comparing the present model to the corresponding one of the second strain gradient elasticity, we observe that they deliver the same dispersion relation for $g^2 = a^2 / 4$, $c = a^2 / 12$, $l^4 = a^4 / 67$, $h^2 = 0$ and $d^4 = b^4 / 20$. However, the present Model VI and the second strain gradient elasticity remain two completely different theories, since the first employs two intrinsic parameters while the second four.

**Model VII:** In this model we consider instead of the nonlocal domains $A : x \in [-a, a]$ and $B : x \in [-b, b]$ the ones represented in Fig. 8(*ii*). More precisely we consider that strains vary quadratically in the domain $A : x \in [-a, a]$ and linearly in the complementary domain $A' : x \in [-\beta, -a] \cup [a, \beta]$. Similarly, velocities vary quadratically in $B : x \in [-b, b]$ and linearly in $B' : x \in [-\gamma, -b] \cup [b, \gamma]$.

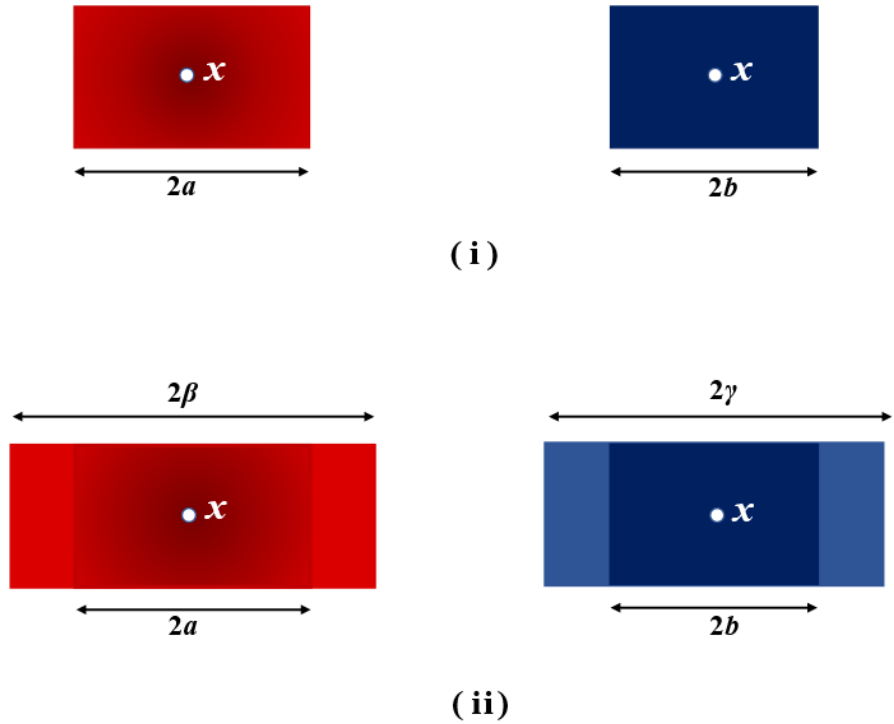

**( i )**

**( ii )**

**Figure 8:** (i) Nonlocal spaces used for the models I, II, III, IV, V and VI. (ii) Nonlocal spaces used for the model VII.

To satisfy the property (26), the kernel $K$ is written for the new nonlocal domain as

$$K\left(\frac{\xi}{\beta}\right) = \frac{1}{1.49364\beta} e^{-\xi^2/\beta^2} \tag{59}$$

and for the evaluation of the potential energy density, we need the following integrals

$$\int_{-a}^{a} \xi^2 K\left(\frac{\xi}{\beta}\right) d\xi = \frac{\beta^2}{1.49364}\left[\frac{1}{2}\sqrt{\pi}\,erf\left(\frac{a}{\beta}\right) - \left(\frac{a}{\beta}\right)e^{-a^2/\beta^2}\right]$$

$$\int_{-a}^{a} \xi^4 K\left(\frac{\xi}{\beta}\right) d\xi = \frac{\beta^4}{1.49364}\left[\frac{3}{4}\sqrt{\pi}\,erf\left(\frac{a}{\beta}\right) - \left(\frac{a^3}{\beta^3} + \frac{3}{2}\frac{a}{\beta}\right)e^{-a^2/\beta^2}\right]$$

(60)

Thus, the potential energy density for the new nonlocal cell has the form

$$\hat{P}(x,t) = \int_{-\beta}^{-a} K\left(\frac{\xi}{\beta}\right)P(x+\xi,t)d\xi + \int_{-a}^{a} K\left(\frac{\xi}{\beta}\right)P(x+\xi,t)d\xi + \int_{a}^{\beta} K\left(\frac{\xi}{\beta}\right)P(x+\xi,t)d\xi$$

$$= \frac{E}{2}\left[\int_{-\beta}^{-a} K\left(\frac{\xi}{\beta}\right)[e(x+\xi,t)]^2 d\xi + \int_{-a}^{a} K\left(\frac{\xi}{\beta}\right)[e(x+\xi,t)]^2 d\xi + \int_{a}^{\beta} K\left(\frac{\xi}{\beta}\right)[e(x+\xi,t)]^2 d\xi\right]$$

$$= \frac{E}{2}\int_{-\beta}^{-a} K\left(\frac{\xi}{\beta}\right)[u' + \xi u'']^2 d\xi$$

$$+ \frac{E}{2}\int_{-a}^{a} K\left(\frac{\xi}{\beta}\right)\left[u' + \xi u'' + \frac{1}{2}\xi^2 u'''\right]^2 d\xi$$

$$+ \frac{E}{2}\int_{a}^{\beta} K\left(\frac{\xi}{\beta}\right)[u' + \xi u'']^2 d\xi$$

(61)

Utilizing relations (60) and after some algebra we eventually obtain

$$\hat{P}(x,t) = \frac{1}{2}E[u']^2 + \frac{1}{2}E\frac{\beta^2}{4}[u'']^2 + \frac{1}{2}E\frac{\beta^2}{4}l_1^2 u' \cdot u''' + \frac{1}{2}E\frac{\beta^4}{16}l_2^4[u'']^2$$

(62)

where

$$l_1^2 = \frac{1}{1.49364}\cdot\left[2\sqrt{\pi}\,erf\left(\frac{a}{\beta}\right) - 4\frac{a}{\beta}e^{-\frac{a^2}{\beta^2}}\right] > 0$$

$$l_2^4 = \frac{1}{1.49364}\left[3\sqrt{\pi}\,erf\left(\frac{a}{\beta}\right) - \left(4\frac{a^3}{\beta^3} + 6\frac{a}{\beta}\right)e^{-\frac{a^2}{\beta^2}}\right] > 0$$

(63)

Expression (62) reproduces (56) for $a/\beta=1$ since $\lim\limits_{a/\beta\to1} l_1^2 = 1$ and $\lim\limits_{a/\beta\to1} l_2^4 = \dfrac{16}{30} = 0.533$

On the other hand, the kinetic energy density reads

$$
\begin{aligned}
\hat{T} &= \frac{1}{2\gamma}\left[\int_{-\gamma}^{-b}\frac{1}{2}\rho[\upsilon(x+\xi,t)]^2\,d\xi + \int_{-b}^{b}\frac{1}{2}\rho[\upsilon(x+\xi,t)]^2\,d\xi + \int_{b}^{\gamma}\frac{1}{2}\rho[\upsilon(x+\xi,t)]^2\,d\xi\right] \\
&= \frac{1}{2\gamma}\left[\frac{\rho}{2}\int_{-\gamma}^{-b}\left[\dot{u}+\xi\dot{u}'\right]^2\,d\xi\right] \\
&\quad + \frac{1}{2\gamma}\left[\frac{\rho}{2}\int_{-b}^{b}\left[\dot{u}+\xi\dot{u}'+\frac{1}{2}\xi^2\dot{u}''\right]^2\,d\xi\right] \\
&\quad + \frac{1}{2\gamma}\left[\frac{\rho}{2}\int_{b}^{\gamma}\left[\dot{u}+\xi\dot{u}'\right]^2\,d\xi\right]
\end{aligned}
\tag{64}
$$

and obtains the final form

$$
\hat{T} = \frac{1}{2}\rho\dot{u}^2 + \frac{1}{2}\rho\frac{\gamma^2}{3}[\dot{u}']^2 + \frac{1}{2}\rho\frac{b^3}{3\gamma}\dot{u}\cdot\dot{u}'' + \frac{1}{2}\rho\frac{b^5}{20\gamma}[\dot{u}'']^2
\tag{65}
$$

The equation of motion corresponding to energy densities (62) and (65) and after the application of Hamilton's principle is

$$
\begin{aligned}
&-\partial_t\left[\frac{\partial L}{\partial\dot{u}}\right] + \partial_{tx}\left[\frac{\partial L}{\partial\dot{u}'}\right] - \partial_{txx}\left[\frac{\partial L}{\partial\dot{u}''}\right] - \partial_x\left[\frac{\partial L}{\partial u'}\right] + \partial_{xx}\left[\frac{\partial L}{\partial u''}\right] - \partial_{xxx}\left[\frac{\partial L}{\partial u'''}\right] = 0 \Rightarrow \\
&E\left[u'' - \frac{\beta^2}{4}\left(1-2l_1^2\right)u'''' + \frac{\beta^4}{16}l_2^4 u''''''\right] - \rho\left(\ddot{u} - \frac{\gamma^3-b^3}{3\gamma}\ddot{u}'' + \frac{b^5}{20\gamma}\ddot{u}''''\right) = 0
\end{aligned}
\tag{66}
$$

In view of (66), it is evident that Model VI reproduces the equation of motion (14) of the second strain gradient elastic theory with the correspondence $g^2 = \beta^2/4$, $C = l_1^2\left(\beta^2/4\right)$, $l^4 = \beta^4/16$, $h^2 = \dfrac{\gamma^3-b^3}{3\gamma}$ and $d^4 = \dfrac{b^5}{20\gamma}$. In the second strain gradient theory both $g^2 - 2C$ and $h^2$ are positive quantities. In the present Model VII, the term $\dfrac{\gamma^3 - b^3}{3\gamma}$ is always positive since $\gamma > b$. On the contrary the term $1 - 2l_1^2$ obtains the following values:

$$1 - 2l_1^2 \begin{cases} > 0, & \text{for } 0 < \dfrac{a}{\beta} < 0.724 \\[2mm] = 0, & \text{for } \dfrac{a}{\beta} = 0.724 \\[2mm] < 0 & \text{for } \dfrac{a}{\beta} = 0.724 \end{cases} \qquad (67)$$

Relation (67) reveals that the Model VII leads to a second strain gradient elastic theory only when the extension of the nonlocal area of potential energy density satisfies the condition $0 < \dfrac{a}{\beta} < 0.724$.

**Conclusions**

In the present work a new one-dimensional nonlocal elastic model is demonstrated and used as generator of all strain gradient elastic models appearing so far in the literature. Comparing to the classical nonlocal elastic theory, the present model appears the following differences:

(i)  Instead of considering nonlocality on stresses, assumes a nonlocal definition of potential energy density defined at a point $x$ of the considered continuum.

(ii) It considers nonlocality in both potential and kinetic energy densities.

(iii) The nonlocality at a point $x$ is confined in a finite region with center the point $x$ and boundary the symmetric points $x$-$a$, $x$+$a$ called horizon of the nonlocality. The horizon of the potential energy density is not necessarily the same with the horizon of the kinetic energy density.

Since only the neighborhood of the point x has a significant input on the nonlocality, strains and velocities were expanded in Taylor power series about $x$. Depending on the truncation of those power series, the following strain gradient models have been derived

**Model I:** When strains and velocities remain constant in the corresponding nonlocal regions, then the classical theory of elasticity is reproduced. Contrary to the models explained below, this is the only model that does not support wave dispersion.

**Model II:** Considering constant strains and linear behavior of velocities, the differential nonlocal model of Eringen (1983) with microinertia effects is obtained. The corresponding internal microinertia parameter is equal to $b^2/3$ with $b$ being the horizon of the kinetic energy density at $x$.

**Model III:** Linear variation of strains and velocities leads to the first strain gradient elastic equation of motion, which has the same form with the corresponding one derived by Polyzos and Fotiadis (2012) via lattice models. The microstructural and microinertia internal length scale parameters have been found to be equal to $a^2/4$ and $b^2/3$, respectively, with $a$, $b$ being the horizon of the potential and kinetic energy density at $x$, respectively. The dynamic behavior of Model III is a result of the competition of two distinguished underline mechanisms, the micro-elasticity and micro-inertia expressed via the intrinsic parameters $a^2/4$ and $b^2/3$, respectively. When $a^2/4=b^2/3$, there is a balance of those two mechanisms and no wave dispersion occurs.

**Model IV:** Introducing in Model III a linear nonlocality of velocities in time, we conclude to a strain gradient elastic theory which proposes an equation of motion with fourth-order derivatives in space and time. The corresponding dispersion curves are similar to those of Model III but the significant difference between Model III and Model IV is that the first employs two internal length scale parameters while the second three.

**Model V & VI:** Keeping more terms in the expansions of strains and velocities than in previous models, we conclude to dispersion relations which can be considered as degenerate cases of the second strain gradient elastic theory. To receive an equation of motion being equivalent to the one of second strain gradient elasticity, an expansion of the previously used horizons is required. This has been done in Model VII.

**Model VII:** Expanding the nonlocal domains according to Fig. 8(ii) and considering the Model VI with linear variation of strains and velocities at the two considered extensions, the second strain gradient equation of motion is reproduced, as this also mentioned in Polyzos and Fotiadis (2012).